\documentclass[pra,aps,twocolumn,showpacs]{revtex4}
\usepackage{times,epsfig,amssymb,amsfonts,amsmath}
\newcommand{\ket}[1]{|#1\rangle}
\newcommand{\bra}[1]{\langle #1|}
\newcommand{\proj}[1]{\ket{#1}\bra{#1}}

\begin{document}

\title{Synthetic spin-orbit coupling in ultracold $\Lambda$-type atoms}
\author{Ming-Yong Ye}
\email{myye@fjnu.edu.cn}
\author{Xiu-Min Lin }
\affiliation{College of Physics and Energy, Fujian Normal University, Fuzhou 350007, China}

\begin{abstract}
We consider the simulation of non-abelian gauge potentials in ultracold atom systems with atom-field
interaction in the $\Lambda$ configuration where two internal states
of an atom are coupled to a third common one with a detuning. We find the simulated non-abelian gauge potentials can
have the same structures as those simulated in the tripod configuration if we parameterize Rabi frequencies properly,
which means we can design spin-orbit coupling simulation schemes based on those proposed in the tripod configuration.
We show the simulated spin-orbit coupling in the $\Lambda$ configuration can only be of a form
similar to $p_{x}\sigma_{y}$ even when the Rabi frequencies are not much smaller than the detuning.
\end{abstract}

\maketitle

\section{introduction}
Many interesting quantum phenomena have been found in condensed
matter physics when electrons have a spin-orbit coupling, such as the spin
Hall effect and the topological insulator \cite{2,3}. Ultracold atom systems
are now regarded as simulation platforms to study condensed matter physics
\cite{1}, therefore it is important to realize spin-orbit couplings in these
systems. It has been shown theoretically that abelian and non-abelian gauge
potentials can be simulated for ultracold atoms via their interaction with
laser fields, and non-abelian gauge potentials can be used to generate
spin-orbit couplings \cite{4,xx,5}. A general spin-orbit coupling for ultracold atoms has not
yet been realized experimentally, however some experimental progress towards
this direction have been made \cite{6,7,8,zhang1,9,10,zhang2}.

Theoretical schemes to simulate spin-orbit couplings for ultracold atoms are
usually proposed with atom-field interaction in the so called tripod
configuration, where two dark states are used to form the effective spin space
\cite{t2,t3,tt1,tt2,tt3,tt4,tt5}. These schemes have a drawback that the two
dark states are not the lowest energy dressed states, hence atom-atom
interactions can induce collisional decay. 
Some authors also consider the simulation of spin-orbit coupling
with atom-field interaction in the $\Lambda$ configuration,
where two lowest energy dressed states are used to form the
effective spin space \cite{L1,L2}. Recently several experiments
have realized the special spin-orbit coupling $p_{x}\sigma_{y}$ for ultracold
atoms via Raman process \cite{9,10,zhang2}, which is a scheme of $\Lambda$ configuration
with Rabi frequencies much smaller than the detuning. Some other kinds of methods
are also proposed to simulate spin-orbit couplings in ultracold atom systems \cite{12,13,you}.

Although many interesting features of ultracold atoms have been theoretically found when they have a general spin-orbit coupling
such as the Rashba and Dresselhaus spin-orbit couplings, currently we can only experimentally achieve
the special spin-orbit coupling $p_{x}\sigma_{y}$ for ultracold atoms via Raman process.
Since Raman process is a scheme of $\Lambda$ configuration
with Rabi frequencies much smaller than the detuning,
it is natural to ask whether more general spin-orbit couplings can be simulated
in $\Lambda$ configuration when Rabi frequencies are not much smaller than the detuning.
We find the answer is NO at least in our concerned $\Lambda$ configuration.

The structure of the paper is as follows. We first give an analytical expression of the simulated non-abelian gauge potentials in our concerned $\Lambda$ configuration,
which can help us to design spin-orbit coupling simulation schemes based on those proposed in the tripod configuration.
We then consider a spin-orbit coupling simulation scheme where two plane waves are used in the $\Lambda$ configuration.
We find the simulated spin-orbit coupling can only be of a form similar to $p_{x}\sigma_{y}$ due to the non-degeneracy of the two lowest energy dressed states.
We also analyze how the relative magnitude of the two lasers affect the simulated spin-orbit coupling Hamiltonian in this scheme.

\section{Non-abelian gauge potential simulation in $\Lambda$ configuration}
A general theory on the simulation of non-abelian gauge potentials for ultracold atoms is presented in Ref. \cite{xx}.
Here we focus on atom-field interaction in the $\Lambda$ configuration.
As shown in FIG. 1, suppose two internal states $\ket{1}$ and $\ket{2}$
of an atom are coupled to a third common one $\ket{3}$ via
laser fields with a detuning. The atom Hamiltonian will be
\begin{equation}
\hat{H}=\frac{{\hat{\vec{p}}}^2}{2m}+\hat{H_0}+V, \label{x1}
\end{equation}
where $\hat{H_0}$ is the atom-field interaction and $V$ is the possible external trapping potential.
In the interaction picture,
\begin{equation}
\hat{H}_{0}=\hbar\Delta\proj{3}+\hbar\left[  \Omega_{1}\ket{3} \bra{1}
+\Omega_{2}\ket{3} \bra{2} +H.c.\right],
\end{equation}
where $\Delta$ is the detuning that is assumed to be positive, $\Omega_{1}$ and $\Omega_{2}$ are Rabi
frequencies. We parameterize two Rabi frequencies as
\begin{equation}
\Omega_{1}=\frac{\Delta}{2}\tan{2\theta}\cos{\phi}e^{iS_1},
\Omega_{2}=\frac{\Delta}{2}\tan{2\theta}\sin{\phi}e^{iS_2},
\end{equation}
with $-\pi/4<\theta<\pi/4$.
Under this parameterization the three eigenstates of $\hat{H}_{0}$ are
\begin{align}
\ket{e_1} &=\sin\phi e^{-iS_{1}}\ket{1} -\cos\phi e^{-iS_{2}}\ket{2} ,\label{x4}\\
\ket{e_2} &=\cos\theta\left(  \cos\phi e^{-iS_{1}}\ket{1} +\sin\phi e^{-iS_{2}}\ket{2}\right)  -\sin\theta\ket{3} ,\nonumber\\
\ket{e_3} &=\sin\theta\left(  \cos\phi e^{-iS_{1}}\ket{1} +\sin\phi e^{-iS_{2}}\ket{2}\right)  +\cos\theta\ket{3} ,\nonumber
\end{align}
with the corresponding eigenvalues being
\begin{equation}
E_{1}=0,E_{2}=-\hbar\Delta\frac{sin^2\theta}{\cos(2\theta)}, E_{3}=\hbar\Delta\frac{cos^2\theta}{\cos(2\theta)}.
\end{equation}
The full quantum state of the atom can be written in the form
$\ket{\Phi}=\sum_{i=1}^{3}\Psi_{i}\ket{e_i}$. Due to the position dependence of the dressed states $\ket{e_i}$,
when we substitute the full quantum state $\ket{\Phi}$ into the Schr$\ddot{o}$dinger equation with $\Hat{H}$ given in Eq. (\ref{x1}),
we can find that the column vector of wave functions $\Psi=(\Psi_1,\Psi_2,\Psi_3)^{T}$ satisfies the Schr$\ddot{o}$dinger equation with Hamiltonian \cite{xx}
\begin{equation}
\tilde{H}=\frac{(\hat{\vec{p}}-\vec{A})^2}{2m}+\tilde{V},
\end{equation}
where $\vec{A}$ and $\tilde{V}$ are $3\times3$ matrices:
\begin{equation}
\vec{A}_{n,m}=i\hbar\bra{e_n} \ket{ \nabla e_{m}},\tilde{V}_{n,m}=E_{n}\delta_{n,m}+\bra{e_n}V\ket{e_m}. \label{x7}
\end{equation}
We note that $\vec{A}$ and $V$ give contribution to the off-diagonal elements and $E_n$ give contribution to the diagonal elements of $\tilde{H}$.

\begin{figure}
\begin{center}
\epsfig{figure=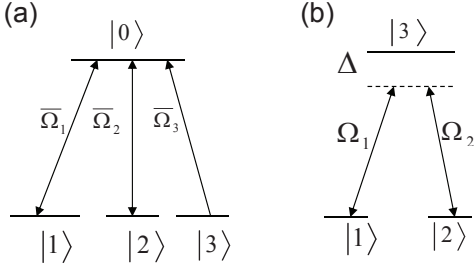,width=0.35\textwidth}
\end{center}
\caption{(a) The tripod configuration and (b) $\Lambda$ configuration.
If in the tripod configuration Rabi frequencies are parameterized as
$\bar{\Omega}_{1}=\bar{\Omega}\sin\theta\cos\phi e^{iS_{1}}$,
$\bar{\Omega}_{2}=\bar{\Omega}\sin\theta\sin\phi e^{iS_{2}}$,
and $\bar{\Omega}_{3}=\bar{\Omega} \cos\theta e^{iS_{3}}$,
and in the $\Lambda$ configuration Rabi frequencies are parameterized as
$\Omega_{1}=\frac{\Delta}{2}\tan{2\theta}\cos{\phi}e^{i(S_1-S_3)}$ and
$\Omega_{2}=\frac{\Delta}{2}\tan{2\theta}\sin{\phi}e^{i(S_2-S_3)}$ with $\Delta$ being the detuning,
then the simulated non-abelian gauge potentials in the $\Lambda$ configuration have exactly the same
structures as those simulated in the tripod configuration.}
\end{figure}
Since $\Delta>0$ and $-\pi/4<\theta<\pi/4$, there are $E_3> E_1 \geq E_2$ and $E_3- E_1\geq\hbar\Delta$.
Now we discuss the condition when the off-diagonal elements of $\tilde{H}$ connecting
 $\Psi_1$ to $\Psi_3$ and $\Psi_2$ to $\Psi_3$ can be neglected, so that it can be used to simulate the movement of a particle with spin-1/2.
Note that $\vec{A}_{n,m}$ usually has a magnitude of momentum $P_L$ of the applied laser fields,
therefore the off-diagonal elements of $\tilde{H}$ connecting $\Psi_1$ to $\Psi_3$ and $\Psi_2$ to $\Psi_3$ can be neglected
when $\frac{P_L^2}{2m}\ll \hbar\Delta$, $|\bra{e_n}V\ket{e_m}|\ll \hbar\Delta$ and atoms move very slowly
(i.e., $\frac{(\vec{p})^2}{2m}\ll \hbar\Delta$).
If these conditions are satisfied the wave functions $(\Psi_1,\Psi_2)^{T}$ will be approximately decoupled from $\Psi_3$ and evolute
under the Hamiltonian \cite{xx}
\begin{equation}
\hat{H}_{eff}=\frac{(\hat{\vec{p}}-\vec{A})^2}{2m}+\tilde{V}+\Phi. \label{x9}
\end{equation}
Here $\vec{A}$, $\tilde{V}$ and $\Phi$ are $2\times2$ matrices, the elements of $\vec{A}$ and $\tilde{V}$
are described in Eq. (\ref{x7}) with $n,m=1,2$, and
\begin{equation}
\Phi_{n,m}=\frac{1}{2m}\vec{A}_{n,3}\cdot \vec{A}_{3,m},~~n,m=1,2.
\end{equation}
The Hamiltonian $\hat{H}_{eff}$ in Eq. (\ref{x9}) simulates the movement of a particle with spin-1/2 in gauge potentials,
where two lowest energy dressed states $\ket{e_1}$ and $\ket{e_2}$ represent spin up and spin down respectively.
Here we emphasize that the $\theta$ is not required to be small to get $\hat{H}_{eff}$, i.e., the magnitudes of Rabi frequencies
$\Omega_{1}$ and $\Omega_{2}$ are not required to be much smaller than the detuning $\Delta$.

We can get an analytical expression of the simulated gauge potentials $\vec{A}$ and $\Phi$ in $\hat{H}_{eff}$
by substituting Eq. (\ref{x4}) into Eq. (\ref{x7}).
But note that the dressed states $\ket{e_1}$, $\ket{e_2}$ and $\ket{e_3}$ have the same mathematical structures
as the two dark states $\ket{D_1}$, $\ket{D_2}$ \cite{xx} and the bright state $\ket{B}=\ket{D_0}$ \cite{14} in the tripod configuration respectively, and in the tripod configuration the simulated gauge potentials can be expressed as
 $\vec{A}_{n,m}=i\hbar\bra{D_n} \ket{ \nabla D_{m}}$ and $\Phi_{n,m}=\frac{1}{2m}\vec{A}_{n,0}\cdot \vec{A}_{0,m}$ \cite{14},
we can conclude immediately that the simulated gauge potentials in our concerned $\Lambda$ configuration have
the same mathematical structures as those simulated in the tripod configuration, i.e., we can obtain
an analytical expression for the simulated gauge potentials $\vec{A}$ and $\Phi$ of $\hat{H}_{eff}$ in our concerned $\Lambda$ configuration
just through replacing $S_{13}$ and $S_{23}$ in
Eq. (13) and Eq. (14) of Ref. \cite{xx} by $S_{1}$ and $S_{2}$ respectively.
Here we write them down for completeness :
\begin{eqnarray}
\vec{A}_{1,1} &=& \hbar(cos^2 \phi \nabla S_2+sin^2 \phi \nabla S_1),  \label{ap} \\
\vec{A}_{1,2} &=& \hbar cos\theta [\frac{1}{2}sin(2\phi) (\nabla S_1-\nabla S_2)-i \nabla \phi],\nonumber \\
\vec{A}_{2,2} &=& \hbar cos^2\theta(cos^2 \phi \nabla S_1+sin^2 \phi \nabla S_2),\nonumber \\
\Phi_{1,1} &=& \frac{\hbar ^2}{2m}sin^2\theta [\frac{1}{4}sin^2(2\phi)(\nabla S_1-\nabla S_2)^2+(\nabla \phi)^2],\nonumber \\
\Phi_{1,2} &=& \frac{\hbar ^2}{2m}sin\theta [\frac{1}{2}sin(2\phi)(\nabla S_1-\nabla S_2)-i\nabla \phi] \nonumber\\
            && \cdot [\frac{1}{2}sin(2\theta)(cos^2 \phi \nabla S_1+sin^2 \phi \nabla S_2)-i\nabla \theta],\nonumber \\
\Phi_{2,2} &=& \frac{\hbar ^2}{2m}[\frac{1}{4}sin^2(2\theta)(cos^2 \phi \nabla S_1+sin^2 \phi \nabla S_2)^2+(\nabla \theta)^2]. \nonumber
\end{eqnarray}
We will give explicit examples of $S_1$ and $S_2$ to show how spin-orbit coupling Hamiltonian can be obtained.

Our parameterization of the two Rabi frequencies not only gives us a convenient way to
obtain an analytical expression of  the simulated gauge potentials as shown above,
but also gives us a way to design spin-orbit simulation schemes in the $\Lambda$ configuration based on those proposed in the tripod configuration.
Suppose there is a scheme to simulate spin-orbit couplings in
the tripod configuration with the Rabi frequencies $\bar{\Omega}_{1}
=\bar{\Omega}\sin\theta\cos\phi e^{iS_{1}}$, $\bar{\Omega}_{2}=\bar{\Omega
}\sin\theta\sin\phi e^{iS_{2}}$ and $\bar{\Omega}_{3}=\bar{\Omega}\cos\theta
e^{iS_{3}}$, then in our concerned $\Lambda$ configuration we can design a counterpart spin-orbit coupling simulation scheme using the
Rabi frequencies $\Omega_{1}=\frac{\Delta}{2}\tan{2\theta}\cos{\phi}e^{i(S_1-S_3)}$ and
$\Omega_{2}=\frac{\Delta}{2}\tan{2\theta}\sin{\phi}e^{i(S_2-S_3)}$ with $\Delta$ being the detuning.
These two schemes simulate the same kind of spin-orbit coupling since the gauge potentials simulated by them
have exactly the same mathematical structures,
where the difference is that there are some restrictions on the parameter $\theta$
and there is a Zeeman term in $\hat{H}_{eff}$ due to the non-degeneracy of the two lowest energy dressed states in the $\Lambda$ configuration.
The known schemes to simulate spin-orbit couplings of Refs. \cite{L1,L2} in the $\Lambda$
configuration can be regarded as the counterpart schemes of Refs. \cite{t2,t3}
in the tripod configuration where standing waves are used.
In the next section we will consider a new spin-orbit coupling simulation scheme
where only two plane waves are used, which can be regarded as
the counterpart scheme of Refs. \cite{tt1,tt2,tt3,tt4,tt5} in the tripod configuration where only plane waves are used.
\section{spin-orbit coupling simulation in $\Lambda$ configuration}
\begin{figure}
\begin{center}
\epsfig{figure=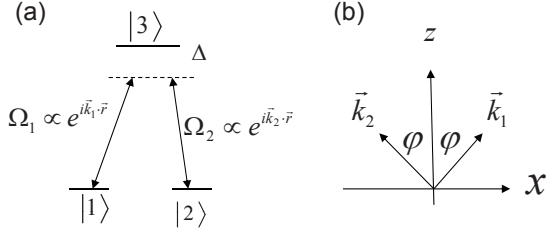,width=0.40\textwidth}
\end{center}
\caption{(a) Two plane waves coupling two lower internal states of an atom to
a common higher internal states.  (b) The directions of the two plane waves.}
\end{figure}
In this section we consider a spin-orbit coupling simulation scheme where two plane waves are used in the $\Lambda$ configuration.
This is an example to show the simulated spin-orbit coupling in the $\Lambda$ configuration can only be of a form
similar to $p_{x}\sigma_{y}$ even when the Rabi frequencies are not much smaller than the detuning.
We also analyze how the relative magnitude of the two lasers affect the form of the simulated spin-orbit coupling Hamiltonian in this scheme.

As shown in FIG. 2 two internal states $\ket{1}$ and $\ket{2}$
of an atom are coupled to a third common one $\ket{3}$ via
two plane waves respectively with the same detuning.
The directions of the two waves are in the x-z plane, and the phases of the two Rabi frequencies are assumed to be
\begin{eqnarray}
 S_1&=&\vec{k}_1 \cdot \vec{r}=kx\sin\varphi+kz\cos\varphi,    \label{sss} \\
 S_2&=&\vec{k}_2 \cdot \vec{r}=-kx\sin\varphi+kz\cos\varphi, \nonumber
\end{eqnarray}
where $k$ is the wave vector of the applied plane waves and $\varphi$ determines the wave directions.
Substitute $S_1$ and $S_2$ into the expressions of simulated gauge potentials $\vec{A}$ and $\Phi$ in Eq. (\ref{ap})
and note that only $S_1$ and $S_2$ are position dependent, we get $A_y=0$ and
\begin{eqnarray}
\frac{A_x}{\hbar k}&=&-\frac{1}{2} sin^2\theta cos(2\phi) sin\varphi \sigma_0+ cos\theta sin(2\phi) sin\varphi\sigma_x   \nonumber \\
                    && -\frac{1}{2}(2-sin^2\theta)cos(2\phi)sin\varphi\sigma_z,  \label{ax}  \\
\frac{A_z}{\hbar k}&=&\frac{1}{2}(2-sin^2\theta)cos\varphi\sigma_0+\frac{1}{2} sin^2\theta cos\varphi\sigma_z, \nonumber
\end{eqnarray}
where $\sigma_x$, $\sigma_y$, $\sigma_z$ are three Pauli matrices and $\sigma_0$ is the $2 \times 2$ identity matrix. If we set $\phi=\pi/4$ and substitute $\vec{A}$ above into $\hat{H}_{eff}$ then we can get a spin-orbit coupling proportional to
$cos\theta sin\varphi p_x \sigma_x +\frac{1}{2} sin^2\theta cos\varphi p_z \sigma_z$ 
which will be proportional $ p_x \sigma_x + p_z \sigma_z$ when we choose proper $\theta$ and $\varphi$ and
 can be further turned into Rashba or Dresselhaus spin-orbit coupling through changing spin basis.
However it will be a different story when the Zeeman term in $\hat{H}_{eff}$ due to the non-degeneracy of the two lowest energy dressed states is considered.

For simplicity we assume the external trapping potential $V=0$, so that the Zeeman term $\tilde{V}$ in the simulated Hamiltonian $\hat{H}_{eff}$ in Eq. (\ref{x9})
will be
\begin{equation}
\tilde{V} =-\hbar\Delta\frac{sin^2\theta}{2\cos(2\theta)}\sigma_0+\hbar\Delta\frac{sin^2\theta}{2\cos(2\theta)}\sigma_z.
\end{equation}
Recall the effective Hamiltonian $\hat{H}_{eff}$ is simulated under the conditions
 $\frac{(\hbar k)^2}{2m}\ll \hbar\Delta$ and $\frac{(\vec{p})^2}{2m}\ll \hbar\Delta$. Now we show
 under these conditions some terms in $\hat{H}_{eff}$ can be further neglected.
 First we find the potential $\Phi$ satisfies
 \begin{equation}
|\Phi_{n,m}|\leq \frac{(\hbar k)^2}{2m}sin^2\theta \ll \hbar\Delta\frac{sin^2\theta}{cos(2\theta)},
\end{equation}
therefore $\Phi$ can be ignored compared to $\tilde{V}$.
Second we write $cos\theta$ in $A_x$ of Eq. (\ref{ax}) as $1-2sin^2(\theta/2)$ and then substitute $A_x$ and $A_z$ into $H_{eff}$,
we can get some energy terms proportional to $sin^2\theta$ or $2sin^2(\theta/2)$ \cite{134}, which are of magnitude
$max[\frac{(\vec{p})^2}{2m},\frac{(\hbar k)^2}{2m}]$ and can also be ignored compared to $\tilde{V}$. Therefore in $\hat{H}_{eff}$ we can safely write $\Phi=0$ and
\begin{eqnarray}
A_x&=&\hbar k sin\varphi[ sin(2\phi) \sigma_x-cos(2\phi) \sigma_z],    \label{y} \\
A_z&=&\hbar k cos\varphi\sigma_0. \nonumber
\end{eqnarray}
We note that this result is obtained due to the non-degeneracy of the dressed states $\ket{e_1}$
and $\ket{e_2}$, not by assuming $\theta$ is close to zero.
From Eq. (\ref{y}) we find that only the movement in the $x$ direction of the atom is coupled to its pseudo spin, and the coupled
movement is governed by
\begin{eqnarray}
\hat{H}_{xs}&=&\frac{(\hat{p}_x-A_x)^2}{2m}+\tilde{V}  \\
            &=&\frac{\hat{p}_x^2}{2m}+2\alpha \hat{p}_x[cos(2\phi) \sigma_z -sin(2\phi) \sigma_x]+h\sigma_z, \nonumber
\end{eqnarray}
where $\alpha=\frac{1}{2m}\hbar k sin\varphi$, $h=\hbar\Delta\frac{sin^2\theta}{2\cos(2\theta)}$ and a constant term
$c=\frac{(\hbar k)^2}{2m}sin^2\varphi-\hbar\Delta\frac{sin^2\theta}{2\cos(2\theta)}$ is neglected.
If we use $\ket{e_1'} = cos\phi\ket{e_1}-sin\phi\ket{e_2}$
and $\ket{e_2'}= sin\phi\ket{e_1}+cos\phi\ket{e_2}$
 instead of $\ket{e_1}$ and $\ket{e_2}$
to represent spin up and spin down respectively, then the coupled Hamiltonian will be
\begin{equation}
 \hat{H'}_{xs}=\frac{\hat{p}_x^2}{2m}+2\alpha \hat{p}_x \sigma_z+h[cos(2\phi)\sigma_z+sin(2\phi)\sigma_x],
\end{equation}
where $2\alpha \hat{p}_x \sigma_z$ represents spin-orbit coupling.
Thus we have shown the simulated spin-orbit coupling in our concerned $\Lambda$ configuration can only
be of a form similar to $p_{x}\sigma_{y}$ even when the Rabi frequencies are not much smaller than the detuning.

The roles of $h$ and $\phi$ seem clear in $\hat{H'}_{xs}$; one controls the magnitude of the "magnetic field" and the other controls its direction.
However, as we change $\phi$, which is determined by the relative magnitude of two lasers, not only $\hat{H'}_{xs}$ but also
the spin basis states $\ket{e'_1}$ and $\ket{e'_2}$ will be changed, and it may be not easy to see the underly physics.
Now we assume $\hbar\Delta$ is so bigger that a small $\theta=\theta_m$ can lead to $h\sigma_z$ dominating $\hat{H}_{xs}$,
then we can study the underly physics only varying $\theta$ between $-\theta_m$ and $\theta_m$.
Since $\theta$ is small, i.e., the Rabi frequencies are much smaller than the detuning,
there is
$\ket{e_2}= \cos\phi e^{-iS_{1}}\ket{1} +\sin\phi e^{-iS_{2}}\ket{2}$ and our simulation scheme reduces to the Raman process in recent experiment \cite{9}. At this time we get $\ket{e_1'} =-e^{-iS_2}\ket{2}$
and $\ket{e_2'} =e^{-iS_1}\ket{1}$, which are independent of $\phi$.
Experimentally we can study the phase transition of $\hat{H'}_{xs}$ due to the change of $\phi$ and $\theta$ as in Refs. \cite{9,10}.
\section{discussion and summary}
We have given an example that the simulated spin-orbit coupling in the $\Lambda$ configuration can only be of a form
similar to $p_{x}\sigma_{y}$ even when the Rabi frequencies are not much smaller than the detuning.
The same conclusion can also be obtained when we consider the spin-orbit coupling simulation schemes in  
Refs. \cite{L1,L2}.
So if we want to get a more general spin-orbit coupling in our concerned $\Lambda$ configuration,
we should find a way to eliminate the Zeeman term in $\hat{H}_{eff}$ due to the 
non-degeneracy of the two lowest energy dressed states. If we assume the trapping potential
$V=V_1\ket{1}\bra{1}+V_2\ket{2}\bra{2}+V_3\ket{3}\bra{3}$ with $V_1=V_2$ and $V_3=V_1-E_2/sin^2\theta$,
then this Zeeman term will be eliminated. However this trapping potential will give a coupling between
$\Psi_2$ and $\Psi_3$ that cannot be ignored.
How to effectively eliminate the Zeeman term due to the
non-degeneracy of the two lowest energy dressed states is still under investigation.

In summary, we have given an analytical expression of the simulated non-abelian gauge 
potentials in our concerned $\Lambda$ configuration based on a special parameterization
of the two Rabi frequencies. We have shown the simulated spin-orbit coupling in our concerned $\Lambda$ configuration can only be of a form
similar to $p_{x}\sigma_{y}$ even when the Rabi frequencies are not much smaller than the detuning.
\section*{ACKNOWLEDGMENTS}
M. Y. Ye would like to thank Z. D. Wang, S. L. Zhu and D. W. Zhang for helpful discussions. This work was supported by the National Natural Science Foundation of China (Grant No. 11004033), the Natural Science Foundation of Fujian Province (Grant No. 2010J01002), and the National Fundamental Research Program of China (Grant No. 2011CBA00203).

\end{document}